\documentstyle[preprint,aps,epsf]{revtex}
\tighten
\begin{document}
\draft
\title{Soliton vacuum energies and the CP(1) model.}
\author{Ian G. Moss}
\address{
Department of Physics, University of Newcastle Upon Tyne, NE1 7RU U.K.
}
\date{April 1999}
\maketitle
\begin{abstract}
The quantum properties of solitons at one loop can be related to phase shifts of waves on the soliton background. These can be combined with heat kernel methods to calculate various parameters. The vacuum energy of a $CP(1)$ soliton in $2+1$ dimensions is calculated as an example.
\end{abstract}
\narrowtext
\section{INTRODUCTION}

Solitons arise in many field theories and their particle-like behaviour has prompted discussions of their quantum properties \cite{coleman}. Recently, some old results of Schwinger \cite{schwinger} have been developed into a useful numerical scheme for calculating the vacuum energy of a soliton at one loop order \cite{graham,farhi}. The aim of this paper is to develop these ideas further using heat kernel methods and discuss some applications. 

The one loop corrections to the soliton energies are likely to be most significant in situations where different classical soliton solutions have the same energy. This happens in an important class of field theories, where the classical solutions fall into topologically seperate families and the solutions in each family saturates an energy bound. The prototype for this behaviour was the BPS monopole solution \cite{bog}, but similar solutions play an important role in superstring theories \cite{sen}. 

The $CP(1)$ model in $2+1$ dimensions is one of the simplest models of this type \cite{ward,din}. The calculation of the vacuum energy of the single soliton solution provides an instructive example of the general technique. A useful comparison can also be made with the $CP(1)$ model in two dimensions, where the one loop contribution to the path integral from instanton solutions has been obtained analytically \cite{berg,fateev}.  

\section{HEAT KERNEL METHOD}

The one loop quantum properties of a soliton solution depend on the normal mode frequencies of classical perturbations about the solution. It will prove convenient to impose boundary conditions at a fixed radius $R$ on the perturbations in order to obtain a discrete spectrum and then take the limit $R\to\infty$.  

The heat kernel is defined by
\begin{equation}
K(t)=\sum e^{-k^2t},
\end{equation}
where the sum extends over the discrete spectrum with values $k^2$. For large $R$, the normal modes with real $k$ approach trigonometric functions of $kR+\phi$, where $\phi$ is a constant phase depending on $k$ and a set of other parameters $\gamma$. If no solitons are present, then the normal mode frequencies (denoted by a superscript zero) are given by
\begin{equation}
k^{(0)}R+\phi^{(0)}=n\pi+\beta,
\end{equation}
where $\beta$ is a phase that depends on the boundary conditions. In the presence of the soliton, the real frequencies are given by
\begin{equation}
kR+\phi=n\pi+\beta.
\end{equation}
In the limit $R\to\infty$, $k\to k^{(0)}$, and the heat kernel can be expressed in terms of the phase shift by  
\begin{equation}
K(t)-K^{(0)}(t)={2\over\pi}\int_0^\infty dk\,
e^{-k^2 t}kt\sum_\gamma\delta_\gamma(k)\label{kernel}
\end{equation}
where $\delta_\gamma=\phi-\phi^{(0)}$.

In fact, the $n=1$ states can disappear from the continuum in the $R\to\infty$ limit if $\delta_\gamma(0)\ne0$ and the derivative $\delta'_\gamma(0)=0$. In this case  the phase shift must be displaced onto a new branch. Furthermore, bound states have their own asymptotic behaviour and need to be added as a seperate contribution to equation (\ref{kernel}). If there are $n_0$ bound states with $k=0$, equation (\ref{kernel}) becomes
\begin{equation}
K(t)-K^{(0)}(t)=n_0+{2\over\pi}\int_0^\infty dk\,
e^{-k^2 t}kt\sum_\gamma(\delta_\gamma(k)-\delta_\gamma(0)).\label{kp}
\end{equation}
Levinson's theorem \cite{levinson} implies that the two expressions (\ref{kernel}) and (\ref{kp}) are usually identical. However, in some situations (including the $CP(1)$ soliton), Levinson's theorem does not apply and equation (\ref{kp}) in the one that must be used.

An important feature of the heat kernel is the well-understood behaviour of the $t\to0$ limit in $d$ dimensions \cite{seeley,gilkey},
\begin{equation}
K(t)\sim t^{-d/2}\sum_{n=0} B_nt^n
\end{equation}
where the heat kernel coefficients $B_n$ determine the one loop ultra-violet divergencies of the theory. Finite parts of physical quantities can be obtained from a regulated heat kernel,
\begin{equation}
K_{\rm reg}(t)=K(t)-t^{-d/2}\sum_{n=0}^{(d+1)/2} B_nt^n
\end{equation}
The heat kernel coefficients are integrals of local polynomials of the background fields and their derivatives. Explicit results are known for the first few coefficients \cite{dewitt,moss}. 

The first term in the asymptotic expansion is equivalent to the free heat kernel $K^{(0)}$. A simple comparison shows that inserting the first Born approximation to the phase shifts $\delta_\gamma^{(1)}$ into equation (\ref{kernel}) produces next term in the asymptotic expansion with the coefficient $B_1$. However, the second order Born approximation $\delta_\gamma^{(2)}$ involves a double integral which contains contributions to $B_2$ and higher terms. The part responsible for $B_2$ can be isolated by taking a derivative expansion and keeping only the leading term. This allows us to express the regulated heat kernel in terms of phase shifts,
\begin{equation}
K_{\rm reg}(t)=n_0+{2\over\pi}\int_0^\infty dk\,
kte^{-k^2 t}\sum_\gamma(\overline\delta_\gamma(k)-
\overline\delta_\gamma(0))\label{rk}
\end{equation}
where $\overline\delta_\gamma=\delta_\gamma-\delta_\gamma^{(1)}-\dots$, subtracting the $n$'th order Born approximations to the phase shift keeping only local terms with $m$ derivatives and $2n+4m\le d+1$.

The heat kernel can be used to find the vacuum energy of a soliton in $d+1$ dimensions or the effective action of an instanton in $d$ dimensions. The vacuum energy of a soliton in $d+1$ dimensions is given by $\case1/2\zeta(-\case1/2)$ \cite{dowker}, where $\zeta(s)$ is the zeta-function
\begin{equation}
\zeta(s)=
{1\over \Gamma(s)}\int_0^\infty dt\, t^{s-1}(K(t)-n_0)
\end{equation}
The effective action of an instanton on $d$ dimensions is given by  $-\case1/2\zeta'(0)$. A proper time cutoff $\epsilon$ on the lower limit of the integral can be introduced to regularise the theory \cite{schwinger2}. Inserting equation (\ref{rk}) gives
\begin{eqnarray}
\zeta(-\case1/2)&=&-{1\over \pi}\int_0^\infty dk
\sum_\gamma(\overline\delta_\gamma(k)-
\overline\delta_\gamma(0))+\hbox{poles}\label{zh}\\
\zeta'(0)&=&{2\over \pi}\int_0^\infty dk\,k^{-1}
\sum_\gamma(\overline\delta_\gamma(k)-
\overline\delta_\gamma(0))+\hbox{poles}
\end{eqnarray}
The finite part is now in a form that can be evaluated numerically and the pole terms, which depend on the heat kernel coefficients, can be absorbed by renormalisation.  

Farhi et al. \cite{farhi} have obtained expressions for the vacuum energy in $3+1$ dimensions by subtracting the full first and second order phase shifts. This corresponds to subtracting an expansion of the heat kernel with nonlocal factors. Such an expansion does indeed exist \cite{barvinski}, although the local version is sufficient for removing the divergencies.

\section{CP(1) SOLITON}

The $CP(1)$ model in $2+1$ dimensions $x^\mu$, $\mu=0 \dots 2$, has a complex field $u$ with Lagrangian density
\begin{equation}
L=-{1\over g^2}{\partial_\mu u\partial^\mu\bar u\over (1+u\bar u)^2}
\end{equation}
where $g^{-2}$ is an energy scale. The solutions to the field equations fall into families characterised by a topological index \cite{ward}. The single soliton solution centered at the origin can be written $u_1=\alpha/z$, where $\alpha$ is a constant and $z$ is the complex coordinate $x^1+ix^2$. 

Perturbations $\xi$ about the soliton satisfy the equation
\begin{equation}
-4D_zD_{\bar z}\xi=k^2\xi
\end{equation}
where $D_{\bar z}=\partial_{\bar z}$ and $D_z=\partial_z-2\bar u_1(\partial_zu_1)/(1+u_1\bar u_1)$. There is one normalisable zero mode $\xi=z^{-2}$ and the remaining spectrum is positive. The first heat kernel coefficient for this operator can be evaluated from standard formulae \cite{dewitt,moss} or by the Atiyah-Singer index theorem \cite{gilkey} to be $B_1=1$.

The perturbation equation separates, and with the decomposition $\xi=(1+\alpha^2/r^2) u_me^{im\theta}$,
\begin{equation}
u_m''+{1\over r}u_m'-{(m+2)^2\over r^2}u_m+k^2u_m=U(r)u_m
\end{equation}
where the potential
\begin{equation}
U(r)=-{4(m+1)\over r^2+\alpha^2}-{8\alpha^2\over (r^2+\alpha^2)^2}
\end{equation}
The boundary conditions are
\begin{equation}
u_m\to\cases{\pm r^{|m+2|}&$r\to0$\cr
A(J_m(kr)-Y_m(kr)\tan\delta_m)&$r\to\infty$\cr}
\end{equation}
The potential does not meet the integrability requirements of Levinson's theorem \cite{levinson} and the numerical solution shows the unusual behaviour that $\delta_m(0)=\pi$ for $m\ge -1$ and $\delta_m(0)=-\pi$ for $m<-2$.

The first Born approximation to the phase shifts can be evaluted 
in terms of Bessel functions, $\delta_m^{(1)}(k)=
\pi(2m-\alpha\partial_\alpha)I_m(k\alpha)K_m(k\alpha)$. The sum over $m$ is not formally convergent, however the limit
\begin{equation}
\lim_{M\to\infty}\sum_{m=-M}^{M}\delta_m^{(0)}(k)=\pi\label{summ}
\end{equation} 
is well defined and when substituted into equation (\ref{kp}) gives the correct conclusion that $B_1=n_0$. This definition of the $m\to\infty$ limit should also be used for actual phase shifts.

The vacuum energy $E$ can be obtained from equation (\ref{zh}), with the observation that the first Born approximation cancels due to (\ref{summ}),
\begin{equation}
E={2\pi\over g^2}-{1\over 2\pi}\int_0^\infty dk\sum_{m=-\infty}^\infty
(\delta_m(k)-\delta_m(0))
\end{equation}
where $2\pi/g^2$ is the classical contribution. The integrand is plotted in figure 1. After performing the integral numerically
\begin{equation}
E={2\pi\over g^2}-0.248\alpha^{-1}
\end{equation}
The energy decreases as the width of the soliton decreases. However, if the width of the soliton changed with time, there would be an adiabatic solution of the form $u=\alpha(t)/z$, but this solution has infinite action. The value of $\alpha$ must therefore remain fixed assymptotically, although it may decrease locally.

The loop expansion parameter for the single soliton is effectively $g^2/\alpha$, and so continuing to analysis to more than one loop would introduce further inverse powers of $\alpha$ \cite{moss2}. This raises the possibility that the energy has a minimum at some value of $\alpha$.

\begin{figure}
\begin{center}
\leavevmode
\epsfxsize=20pc
\epsffile{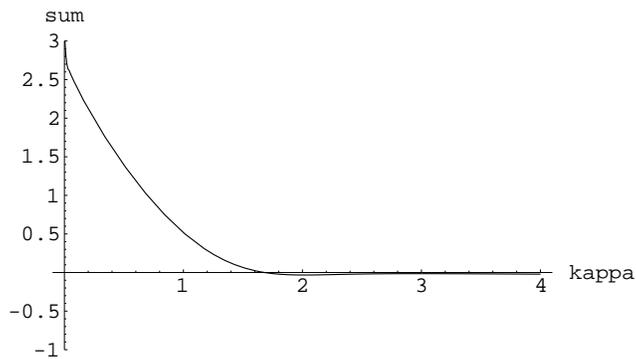}
\end{center}
\caption{The integrand for the vacuum energy of the $CP(1)$ soliton plotted as a function of $\kappa=k\alpha$, with angular modes up to $M=36$. (The cusp at $\kappa=0$ comes from the $m=-1$ mode and, since $\delta_{-1}'(0)\ne0$, $\delta_{-1}(0)$ has not been subtracted.)
\label{fig1}}
\end{figure}

\end{document}